\documentstyle[prl,aps,psfig]{revtex}
\begin{document}
\draft
\twocolumn[\hsize\textwidth\columnwidth\hsize\csname
@twocolumnfalse\endcsname

\widetext
\title{Stripes and spin-incommensurabilities are favored  by lattice 
anisotropies}
\author{Federico Becca,$^{1}$ Luca Capriotti,$^{2}$ and Sandro Sorella$^{3}$}
\address{
${^1}$  Institut de Physique Th\'eorique, Universit\'e de Lausanne, CH-1015 Lausanne, Switzerland\\
${^2}$ Istituto Nazionale per la Fisica della Materia, Unit\`a di Firenze, I-50125 Firenze, Italy \\
${^3}$ Istituto Nazionale per la Fisica della Materia and International School for Advanced Studies, 34013 Trieste, Italy\\
}
\date{\today}
\maketitle
\begin{abstract}
Structural distortions in cuprate materials give a natural origin 
for anisotropies in electron properties. We study a modified one-band
$t{-}J$ model in which we allow for different hoppings and 
antiferromagnetic couplings in the two spatial directions
($t_x \ne t_y$ and $J_x \ne J_y$).
Incommensurate peaks in the spin structure factor show up only in the presence 
of a lattice anisotropy, whereas charge correlations, indicating
enhanced fluctuations at incommensurate wave vectors, are almost 
unaffected with respect to the isotropic case.
\end{abstract}
\pacs{74.20.Mn, 71.10.Fd, 71.10.Pm, 71.27.+a}
]
\begin{figure}
\centerline{\psfig{bbllx=-10pt,bblly=-70pt,bburx=770pt,bbury=550pt,%
figure=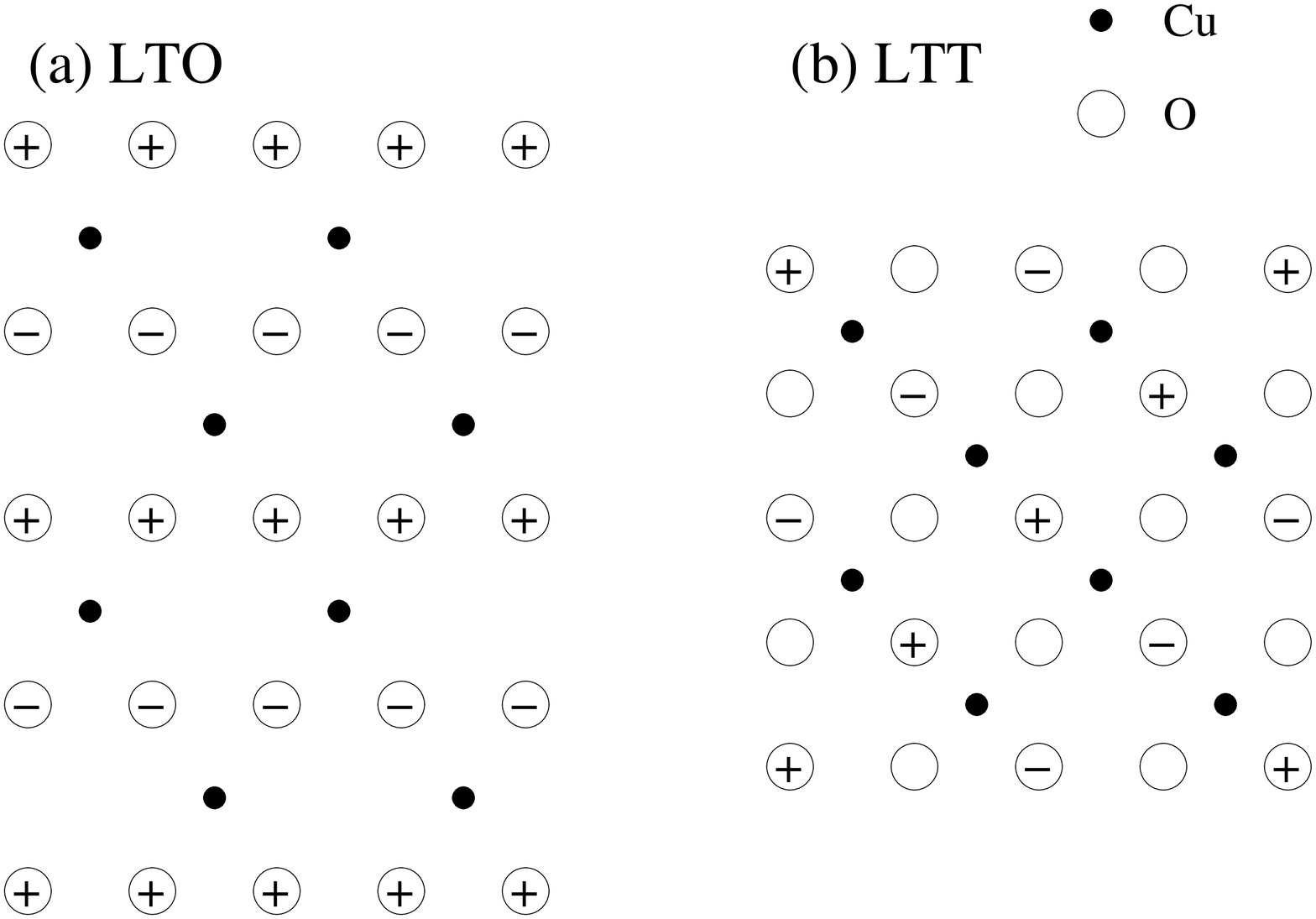,width=60mm,angle=0}}
\caption{\baselineskip .185in \label{lttlto}
A schematic representation of the ${\rm CuO_{2}}$ plane in the LTO (a) 
and in the LTT (b) lattice structure. The black circles represent
Coppers, which are in the plane in both structures. Large
circles represent Oxygens, the symbol $+$ ($-$) indicates that the atom is
shifted above (below) the Copper plane.}
\end{figure}

\narrowtext
The experimental finding of electron inhomogeneities, the so-called
stripes, in doped Mott insulators has recently stimulated a great
debate. The first experimental detection of stripes in a family of 
${\rm La_{2-x}Sr_{x}CuO_{4}}$ (LSCO) cuprates was 
achieved in a Neodymium co-doped compound,
${\rm La_{1.6-x}Nd_{0.4}Sr_{x}CuO_{4}}$.  By using neutron diffraction
measurements,
Tranquada and co-workers,\cite{tranquada1,tranquada2} 
found that the magnetic peak at $Q=(\pi,\pi)$ splits
by a quantity ${\rm x}$, giving rise to four incommensurate peaks around $Q$.
Moreover, the Bragg peaks split by $2{\rm x}$ around the $\Gamma$ point. 
This fact has been interpreted with the formation of
charge domain walls separated by antiferromagnetic regions where
no holes are present. Moreover, the incommensurate magnetic peak 
reveals a $\pi$-phase shift of the staggered magnetization when crossing 
the domain walls.
Further studies \cite{mook} show that similar 
low-energy magnetic peaks occur also in 
${\rm YBa_{2}Cu_{3}O_{6.6}}$, indicating that the incommensurate spin
peaks are common features of all cuprates.
The ${\rm Nd}$-doping on LSCO produces a  structural transition of
the ${\rm CuO_2}$ plane from the low-temperature orthorhombic (LTO) to 
the low-temperature tetragonal (LTT) phase.\cite{axe} 
In the LTO phase the Oxygen atoms are displaced out the Copper plane
and there is only one ${\rm Cu{-}O}$ bond length, see Fig.~\ref{lttlto}(a). 
In the LTT phase, instead, the 
${\rm Cu{-}O}$ bonds are not equivalent in the two directions. 
In one direction the Oxygen atoms are precisely in the Copper plane,
while in the other the 
Oxygens are displaced out of the Copper plane, see Fig.~\ref{lttlto}(b).
The ${\rm Cu{-}Cu}$ hopping depends on the ${\rm Cu{-}O}$
bond and it is isotropic in the LTO phase and 
anisotropic in the LTT one. 

In this letter, we show that the anisotropies in the ${\rm CuO_2}$ 
may help in determining incommensurate electron
correlations.\cite{normand,white} 
In particular, the anisotropy in the antiferromagnetic
coupling favors huge incommensurate spin peaks. These peaks, although
reduced, persist also when considering the anisotropy in the hopping,
indicating that the structural transition may be responsible for the
electron incommensurability, even if other physical effects may
cooperate in the occurrence of the striped phase.
The fact that no sign of 
enhanced charge fluctuations are detected in the presence of spatial
anisotropies does not contradict the experimental finding, which
are sensible to the spin degrees of freedom and not to the electronic charge. 
Finally, we show that the superconducting pairing is suppressed only 
along the stripe, supporting a coexistence of
incommensurate spin fluctuations and superconductivity, as found in
Ref.~\cite{tranquada2}.

We consider a two-dimensional $t{-}J$ model with different
coupling in the $x$ and $y$ directions:
\begin{eqnarray}\label{tj}
{\cal H} &=&
\sum_{i,\mu=x,y} J_{\mu} \left ( {\bf S}_i \cdot {\bf S}_{i+\mu} -
\frac{1}{4} n_i n_{i+\mu} \right ) \nonumber \\
&-& \sum_{\sigma,i,\mu=x,y} t_{\mu}
{\tilde c}^{\dag}_{i,\sigma} {\tilde c}_{i+\mu,\sigma} +H.c.,
\end{eqnarray}
where ${\tilde c}^{\dag}_{i,\sigma}=c^{\dag}_{i,\sigma}
\left ( 1- n_{i,\bar \sigma} \right )$, 
$n_i$ and ${\bf S}_i$ are the electron density and spin on site $i$, 
respectively. The anisotropies are given by the fact that
$t_\mu=\alpha_\mu t$ and $J_\mu=\beta_\mu J$.

We use different Quantum Monte Carlo (QMC) techniques to estimate the
ground-state properties of the Hamiltonian (\ref{tj}).
In particular, we apply the recent QMC method based
on the application of $p$ Lanczos steps to a given variational 
wave function.\cite{sorella}  Moreover, we also used
the fixed-node \cite{ceperley} and stochastic 
reconfiguration (SR)\cite{sorella} approximations.
We consider the $p=0$ and the $p=1$ state as the guiding
wave function for the fixed-node method, and, hereafter, the symbols FN 
and FNLS indicate the fixed-node approximation with the $p=0$ and $p=1$
wave functions, respectively. 
Although these techniques provide a considerable improvement of the 
simple variational calculation, 
they crucially depend on the choice of the guiding wave function.
Both the fixed-node and the SR are exact both in the limit 
of strong anisotropy ($t_x=J_x=0$ and any doping) and in the low-doping 
limit (any anisotropy) and therefore they represent reliable 
approximations for this problem.

The best variational wave function is the projected BCS state
\begin{equation}\label{wf}
|\Psi_{G} \rangle = {\cal P}_N {\cal P}_G {\cal J} {\rm exp}
\left ( \sum_{i,j}
f_{i,j} c^{\dag}_{i,\uparrow} c^{\dag}_{j,\downarrow} \right )
|0 \rangle,
\end{equation}
where ${\cal P}_N$ is the projector onto the subspace of $N$ particles,
${\cal P}_G$ is the Gutzwiller projector, which forbids doubly
occupied sites, ${\cal J} ={\rm exp}(\sum_{i,j} v_{i,j} h_i h_j )$
is a Jastrow factor, defined in term of the hole density at site $i$ 
$h_i= (1-n_{i\uparrow}) (1-n_{i,\downarrow})$, and $v_{i,j}$ are 
variational parameters.
The variational parameters $f_{i,j}$ represent the pair amplitude of
the BCS wave function.

The wave function (\ref{wf}) naturally describes a Resonating
Valence Bond (RVB) state in which preformed pairs
become superconducting by doping.\cite{anderson,randeria}
Whenever the lattice does not
break any translational and rotational symmetries and 
$J \lesssim 0.5t$, the wave function (\ref{wf}) is an exceptionally 
good approximation of the  ground state of the $t{-}J$ model.\cite{sorella} 
For instance, in the case of $4$ holes on a 
$26$ lattice and $J=0.5t$, the overlap of this wave function with the
exact ground state is $0.88$.
In the isotropic model, by using QMC and starting from this 
very accurate wave function,
it is then possible to obtain almost exact ground-state properties
also for large lattices, indicating a small but finite superconducting
order parameter and no charge inhomogeneities.\cite{calandra}
On the other hand, density matrix renormalization group (DMRG) calculations 
for $J \lesssim 0.5t$ are
interpreted as a strong tendency of the ground state toward the
formation of static stripes.\cite{dmrg} These results, however,
were obtained for rectangular clusters and a particular choice of 
the boundary conditions: periodic (open) along the short (long) direction.
The first choice breaks the rotational symmetry, the second one breaks also
the translational invariance. 

\begin{figure}
\centerline{\psfig{bbllx=50pt,bblly=230pt,bburx=500pt,bbury=670pt,%
figure=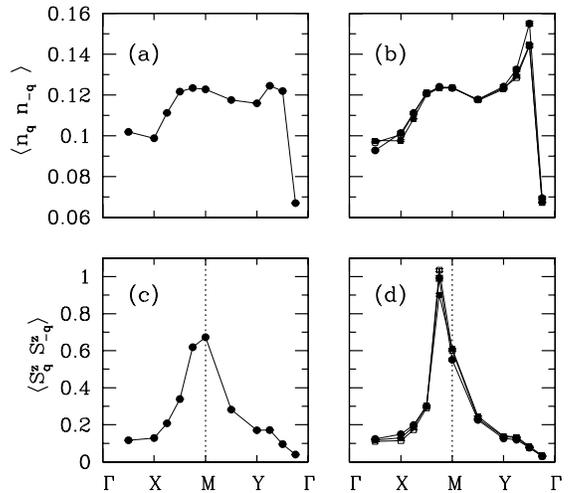,width=70mm,angle=0}}
\caption{\baselineskip .185in \label{8x4}
$N(q)$ and $S(q)$ for $4$ holes on a $4 \times 8$ lattice and $J=0.4t$.
(a) Variational with $p=0$, (b) FN (empty squares), 
FNLS (full squares) and SR (full circles) results for $N(q)$.
(c) Variational with $p=0$, (d) FN (empty squares), 
FNLS (full squares) and SR (full circles) results for $S(q)$.
$\Gamma = (0,0)$, $M = (\pi,\pi)$, $X = (\pi,0)$, $Y = (0,\pi)$.}
\end{figure}

In order to understand how stripes are stabilized, 
it is important to release in the model only the rotational invariance, 
without spoiling the translational symmetry.
In fact the LTO $\rightarrow$ LTT transition leads to a
conformation which naturally breaks the rotational invariance, 
preserving the translational symmetry of the underlying lattice.
We begin by considering a rectangular lattice 
$4 \times 8$ with periodic boundary conditions (PBC) on both directions
and $4$ holes, for
the case with no anisotropy in the couplings, $\alpha_\mu=\beta_\mu=1$ and
$J=0.4t$. In Fig.~\ref{8x4}(a,b), we report the density-density correlations
$N(q)=\langle n_{q} n_{-q}\rangle$ for the variational wave function
with $p=0$,
for the FN, FNLS and SR approximations. In this case the variational state 
does not
present any particular structure in the density correlations, and the
$p=1$ and $2$ results do not change the overall nature of the state.
Instead, the more involved FN, FNLS and SR approaches give rise to a 
huge peak at $q_c=(0,\pi/2)$.
The FN, FNLS and SR methods are able to detect the right
tendency of the charge arrangement, because the approximate ground state 
sampled by these techniques can allow the formation of many-holes 
bound states, not present in the wave function (\ref{wf}),
containing only two-body correlations.

From the experimental point of view, the most important signature of
stripes is the appearance of well-defined incommensurate peaks in 
the magnetic structure factor $S(q,\omega)$ at small energies.
It is really impressive that, by using the FN, FNLS and SR
approaches, we find the same fingerprint of stripes in the equal-time 
correlations $S(q)=\langle S^z_{q} S^z_{-q} \rangle$,
as reported in Fig.~\ref{8x4}(c,d). 
For the most accurate calculations, the $Q$-peak splits into two peaks
at $q_s=(\pi, \pi \pm \pi/4)$, indicating that the 
modulation in the spins is twice the one in the charge.\cite{gazza}

This calculation clearly shows that stripes can be derived   
by weak perturbations over the state of Eq.~(\ref{wf}):
many pairs may bound together 
along a preferred direction, and are stabilized in the $t{-}J$
model, if the rotational symmetry of the ${\rm CuO_{2}}$ 
plane is explicitly broken.

\begin{figure}
\centerline{\psfig{bbllx=30pt,bblly=250pt,bburx=510pt,bbury=660pt,%
figure=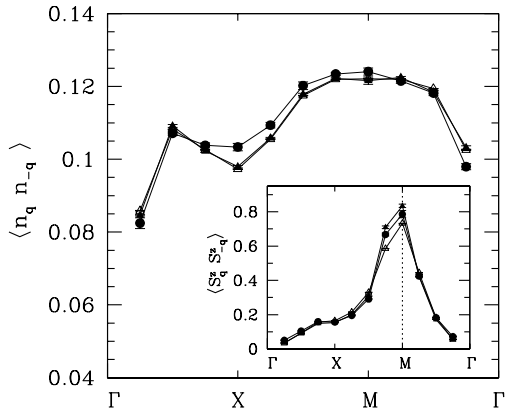,width=70mm,angle=0}}
\caption{\baselineskip .185in \label{8x8}
$N(q)$ for $8$ holes on a $8 \times 8$ isotropic lattice and $J=0.4t$. 
Variational with $p=0$ (empty triangles), FN (full triangles) and
SR (full circles) results are reported. In the inset: the same for $S(q)$.}
\end{figure}

In the following we compare square lattices with PBC
on both directions with and without anisotropies.
In the latter case, for moderate hole doping $\delta \sim 0.1$ and 
$J \sim 0.5t$, the wave function (\ref{wf}) is an accurate
approximation of the ground state:
as shown in Fig.~\ref{8x8} there are no changes in the 
correlation functions by improving the approximation used. 
As emphasized in Figs.~\ref{8x4} and \ref{8x8}, within our approach,
there is a strong influence of the boundary conditions for the 
stabilization of stripes. 
Within the DMRG, it was not possible to realize the important
role played by the boundary conditions.
Indeed, it is well known that DMRG calculations are not accurate for PBC,
especially on two-dimensional clusters,
where it is not possible to reproduce the homogeneous ground state even 
on the $6 \times 6$.
Although for few chains our approach is less accurate -- but 
qualitatively correct -- than DMRG,\cite{dagottone}
it is more suitable for two-dimensional lattices.
The fact that QMC is competitive with DMRG is confirmed by the fact that
on an $8 \times 8$ lattice with open boundary conditions (OBC) 
on both directions (where the DMRG is much more accurate than PBC), 
$J=0.4t$ and $8$ holes, the SR variational energy, 
$E=-39.296(6)t$, is lower than a ``state of the art'' (4200 states kept) DMRG
calculation, $E=-39.2503t$, and very close to the extrapolated result,
consistent for both techniques, $E=-39.44(4)t$.\cite{private} 

\begin{figure}
\centerline{\psfig{bbllx=30pt,bblly=200pt,bburx=550pt,bbury=710pt,%
figure=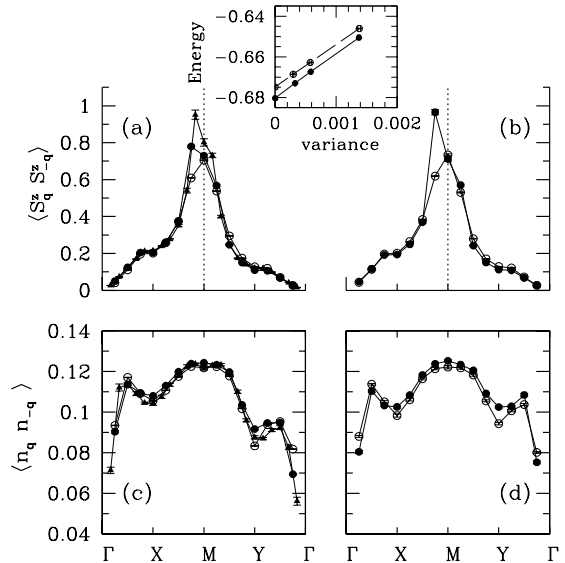,width=80mm,angle=0}}
\caption{\baselineskip .185in \label{8x8anis}
(a): S(q) for $\alpha_x=1.1$, $\alpha_y=0.9$ and 
$\beta_\mu=\alpha_\mu^2$ (case 1). Empty and full circles: variational 
with $p=0$ and SR results for $8$ holes on a 
$8 \times 8$ lattice. Full triangles: SR results for $18$ holes on
a $12 \times 12$ lattice.
(b): S(q) for $8$ holes on the $8 \times 8$ lattice with 
$\alpha_\mu=1$, $\beta_x=1.2$ and $\beta_y=0.8$ (case 2).
(c): the same as (a) for $N(q)$. (d): the same as (b) for $N(q)$. 
In the inset: the energy per site extrapolation for 
$p=0,1,2$ Lanczos steps for the case 1 (continuous line)
and for the case 2 (dashed line), for the $8 \times 8$
lattice. The extrapolated points are also shown.}
\end{figure}

The presence of anisotropies strongly affects the outcome given by 
Fig.~\ref{8x8}.
Indeed, by comparing Fig.~\ref{8x8anis} with the analogous one without 
anisotropy (Fig.~\ref{8x8}), it appears that 
the spin-spin correlations are strongly affected by the 
anisotropy: though at the pure $p=0$ variational level $S(q)$ has a 
broad peak around the antiferromagnetic vector $Q$, within 
the SR technique, incommensurate peaks at $(\pi,\pi \pm \pi/4)$ show up, see
Fig.~\ref{8x8anis}(a,b). We expect that the exact value of these peaks
is underestimated and that they are much more pronounced in the exact 
ground state. Indeed, whenever the ground-state  
correlations  are qualitatively different from the ones
of the projected BCS wave function, by using the SR (or the 
fixed-node) approach it is possible to detect the most
relevant changes in the correlations. 
The incommensurate peaks for $8$ holes on $8 \times 8$ and $18$ holes on 
$12 \times 12$ lattices
are consistent with two half-filled and $3/4$-filled stripes, respectively.
It is worth noting that $3$ half-filled stripes in the $12 \times 12$ give 
rise to $3$ $\pi$-shifts which are not compatible with PBC.
Thus, it is not clear at present what the is most stable stripe filling 
in the thermodynamical limit 
(from half to completely filled stripes\cite{white2}). 

\begin{figure}
\centerline{\psfig{bbllx=20pt,bblly=0pt,bburx=280pt,bbury=200pt,%
figure=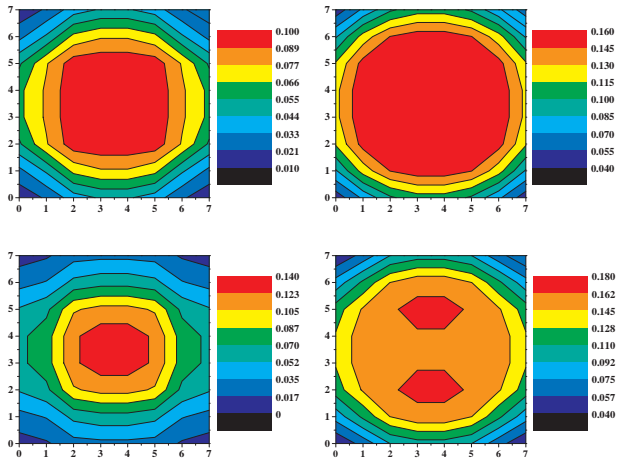,width=90mm,angle=0}}
\caption{\baselineskip .185in \label{open}
On the left: 
average hole density profile for $4$ holes in a $8 \times 8$ lattice 
with OBC and $J=0.4t$, $\alpha_\mu=1$, 
$\beta_x=1.2$ and $\beta_y=0.8$ (case 2 of Fig.~\ref{8x8anis}),
for the $p=0$ calculation (upper-left panel) and for the SR technique
(lower-left panel). On the right: the same for $8$ holes.}
\end{figure}

The effect of the lattice anisotropy on the charge correlations is 
much less evident and $N(q)$ does not show sizable differences between 
different approaches, see Fig.~\ref{8x8anis}(c,d).
We stress that the neutron scattering is only sensible to the electronic
spin and there is no direct information on electronic charge.
The finite peak at small-$q$ in the $N(q)$ may be attributed to dynamical 
fluctuations of the stripes,
which do not affect the $S(q)$ incommensurate peak, but destroy a 
coherent response in the static charge correlations.
In order to have this outcome, it is important that the spins across the 
stripes are strongly antiferromagnetically correlated.\cite{martins}
We have verified that, in the anisotropic case, the spin-spin 
correlation across an hole has a strong antiferromagnetic character.
The above scenario is confirmed by the fact that the anisotropy in 
$t$ favors the transverse motion of the
domain walls, yielding a suppression of the incommensurate peak of $S(q)$,
as shown in Fig.~\ref{8x8anis}(a,b) for the $8 \times 8$ case.
Well defined static stripes are instead defined in the system by 
considering OBC.
As shown in Fig.~\ref{open}, in the $8 \times 8$ cluster there 
is a clear tendency to formation 
half-filled stripes, although in the variational calculation 
no signature of inhomogeneities is present. For $4$ holes there is
a single stripe at the center of the lattice, whereas for
$8$ holes, two of such features clearly appear, indicating the half-filled
character.

\begin{figure}
\centerline{\psfig{bbllx=20pt,bblly=200pt,bburx=560pt,bbury=670pt,%
figure=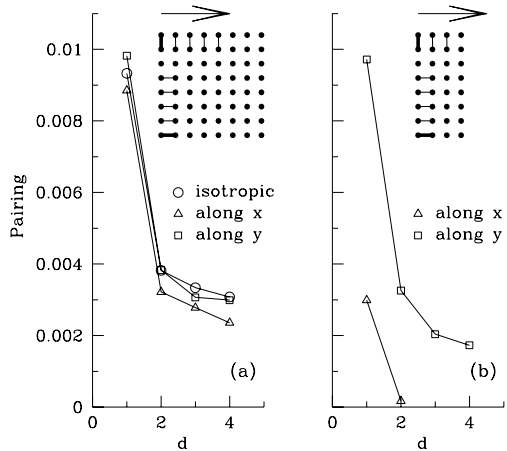,width=70mm,angle=0}}
\caption{\baselineskip .185in \label{pairing}
(a): pairing correlation function $\Delta_{i,j,k,l}$ obtained with SR 
for the isotropic $8 \times 8$ lattice of Fig.~\ref{8x8} (circles) and the 
anisotropic one of Fig.~\ref{8x8anis}(b,d) (triangles and squares).
(b): the same for the $4 \times 8$ lattice of Fig.~\ref{8x4}.
The pairs $(k,l)$ (thin lines) are obtained by moving the pair $(i,j)$ 
(thick line) parallel to the $x$ or the $y$ axis, $d$ is the distance
between pairs. The direction of the stripes is defined by the arrow.}
\end{figure}

In order to study the role of the anisotropies on the 
superconductivity, we consider the pair-pair correlation function
$\Delta_{i,j,k,l}$, which creates a singlet in the sites $i$ and $j$ and
destroys it in the sites $k$ and $l$. The results are reported in 
Fig.~\ref{pairing}.
The effect of the anisotropy is to depress the pairing function
along the stripe, whereas the pairing remains almost unchanged in the 
perpendicular direction.
This effect is particularly strong in the $4 \times 8$ lattice where 
pairing correlations are suppressed by more than an order of 
magnitude along the stripe.

In conclusions, in the presence of anisotropies, our finding is consistent 
with fluctuating stripes, where the 
$\pi$-phase shift gives rise to incommensurate peaks in the $S(q)$
and to an almost featureless $N(q)$.
We finally remark that stripes do not necessarily suppress 
superconductivity, but large pairing 
correlations can be obtained in the direction perpendicular to the 
stripes.
On the other hand, $d$-wave superconductivity is a very robust property of an 
isotropic doped antiferromagnet, implying that 
the RVB scenario represents 
a reasonable explanation of high-temperature superconductivity.
 
We thank E. Dagotto, C. Gazza, G.B. Martins, T.M. Rice,
M. Calandra, A. Parola, M. Fabrizio, 
A. Bianconi and C. Morais Smith for useful comments.
This work has been partially supported by MURST (COFIN99).


\end{document}